# Preventing Spoliation of Evidence with Blockchain: A Perspective from South Asia

Ali Shahaab*
Cardiff School of Technologies,
Cardiff Metropolitan University,
Cardiff, CF5 2YB, UK
ashahaab@cardiffmet.ac.uk

Chaminda Hewage
Cardiff School of Technologies,
Cardiff Metropolitan University,
Cardiff, CF5 2YB, UK
chewage@cardiffmet.ac.uk

Imtiaz Khan
Cardiff School of Technologies,
Cardiff Metropolitan University,
Cardiff, CF5 2YB, UK
ikhan@cardiffmet.ac.uk

## ABSTRACT

Evidence destruction and tempering is a time-tested tactic to protect the powerful perpetrators, criminals, and corrupt officials. Countries where law enforcing institutions and judicial system can be comprised, and evidence destroyed or tampered, ordinary citizens feel disengaged with the investigation or prosecution process, and in some instances, intimidated due to the vulnerability to exposure and retribution. Using Distributed Ledger Technologies (DLT), such as blockchain, as the underpinning technology, here we propose a conceptual model – '*EvidenceChain*', through which citizens can anonymously upload digital evidence, having assurance that the integrity of the evidence will be preserved in an immutable and indestructible manner. Person uploading the evidence can anonymously share it with investigating authorities or openly with public, if coerced by the perpetrators or authorities. Transferring the ownership of evidence from authority to ordinary citizen, and custodianship of evidence from susceptible centralized repository to an immutable and indestructible distributed repository, can cause a paradigm shift of power that not only can minimize spoliation of evidence but human rights abuse too. Here the conceptual model was theoretically tested against some high-profile spoliation of evidence cases from four South Asian developing countries that often rank high in global corruption index and low in human rights index.

## CCS CONCEPTS

• **Security and privacy** → Database and storage security; Data anonymization and sanitization;

## KEYWORDS

blockchain technology for evidence preservation, distributed ledger technologies, evidence protection, spoliation of evidence in developing countries

## 1 INTRODUCTION

Spoliation of evidence refers to altering, destruction, tampering, fabricating or withholding evidence relevant to a legal proceeding [1]. It is also known that spoliation of evidence is a common phenomenon across the globe to protect the perpetrators, corrupt officials and politicians [2-7] For many developing countries spoliation of evidence is purposefully done to cover up corruption or to protect powerful individuals. Since the judicial system in these countries is predominantly paper base, fire is usually used to destroy critical evidence, along with wide range of techniques to tamper and fabricate documents [8]. Furthermore, bribery and intimidation are commonly used to withhold evidence [8]. Paper based evidence archived in a central repository under the custodianship of a central authority are thought to be the main vulnerable points of spoliation of evidence, since in most of the spoliation cases, these repositories or authorities were targeted.

With growing accessibility of smart phones in developing countries, ordinary citizens are becoming well versed to capture critical evidence. This is evident from the growing trend of using digital evidence (photos, emails, audio-video recordings etc.) in investigations and court proceedings. However, several outdated laws and regulations may consider these digital trails of events to be inadmissible in the court of law, i.e. laws requiring first-hand witness to provide evidence [9]. It is not hard to imagine that with the growing number of trades, transactions, and communication in virtual world, paper-based evidence can soon be replaced by digital evidence. For example, a viral video on social media may be a reason for the authorities to take action and provision of justice. However, despite this imminent and inevitable transformation, spoliation of evidence will continue or may increase with digital evidence in near future unless the central aspect of archiving and administration is changed. Furthermore, this transformation will require the legislative bodies to update the laws and make room for the new dimension of evidence collection and submission.

Addressing this paradox, this paper proposes *EvidenceChain*, a hybrid blockchain model to protect vital evidence from spoliation. Using Ethereum [10] blockchain network, which has over 25K nodes across the globe and a peer-to-peer distributed file system, IPFS [11], this conceptual model aims to facilitate evidence collection from general public while preserving their privacy, and secure archiving of evidence in a distributed, synchronized and temper evident manner. In the proposed model, ordinary citizens can upload evidence without compromising their privacy, using anonymous communication facilities of web services like Tor [12]. The submitted evidence is then protected from any malicious tempering attempts using the cryptographic features and distributed nature of blockchain technology. Bypassing authorities to produce evidence will facilitate confidence in public and enhance the usability of the system, which in the long run will encourage citizens to come forward with vital evidence, who otherwise would feel vulnerable. Distributed aspect of the blockchain which made it indestructible against single point of attack or compromise, will provide security, and immutability aspect will protect against tampering or fabrication.

From the authorities perspective the model will use private-permissioned blockchain network [13] which can be accessed only by authorized personnel. This private blockchain will be linked with public blockchain as a tiered architecture model, allowing users to communicate selectively. Through this hybrid blockchain interface, evidence related to the investigation will be made available to the authorities for authenticity and relevancy assessment purpose. Upon validation, the evidence will be classified as authentic and preserved using decentralized storage system (IPFS) and the proof of integrity of the evidence will be preserved on the immutable blockchain. Consensus based vetting in the private tier will safeguard against any kind of biasness from authority side while simultaneously facilitate to filter out false evidence submitted by the general public.

In order to illustrate why and how such model can be used in South Asian context, this paper outline the breadth and depth of spoliation of evidence in South Asian countries in section 2. Blockchain and smart contract technologies are described in section 3, following with examples of related work in section 4 and the details of *EvidenceChain* in section 5. The proposed model is then evaluated in section 6 and limitations and future work are discussed in section 7 and 8, respectively. The paper finally concludes in section 9.

## 2 SPOLIATION OF EVIDENCE IN SOUTH ASIA

In South Asian countries, where most of the evidences are still paper and physical witness based, different types of spoliation of evidence occur. Examples of some high-profile spoliation of evidence are listed below under three major types of spoliation of evidence.

### 2.1 Destruction of Evidence

Fire is the most common form of evidence spoliation tactics in South Asian countries. In 2004, during the investigation of collusion between land mafia and authorities, fire broke out at Lahore Development Authority (LDA) and number of files associated to that case related plots of land went missing [14]. Similar incident also occurred during 2013 but this time eight lives were also lost, along with files containing information worth of PKRs 1.5 trillion (approximately USD 9.27 billion) [15]. In 2018 fire broke out in Income Tax Department office in Mumbai, India during the Rs 14,000 crore banking fraud case, allegedly committed by absconding jeweler Nirav Modi and his uncle Mehul Choksi during 2011-2017 period [16]. In 2019, Bangladesh Standards and Testing Institution (BSTI), a government authority, itself destroyed all the evidence relating to the prosecution of 61 companies accused of marketing 73 uneatable packaged food items and deliberately framed innocent individuals [17].

### 2.2 Tampering/Fabrication of Evidence

In a recent incident in Sri Lanka, phone call recordings of MP Ranjan Ramanayake were leaked online [18] as an evidence of collusion between Mr. Ranjan and high profile government officials including high court judges with regard to alleged corruption cases. However, some of these conversations available on social media seems to be fabricated, which caused doubt about the previously leaked conversations among the general public [19]. The joint investigation team (JIT) probing the killing of four people by the Counter-Terrorism Department (CTD) personnel in an alleged encounter in Sahiwal, Pakistan, in 2019, found evidence of tampering within the digital records by the higher-ups authorities [20]. In another incident of evidence tampering, investigating officials for the assassination case of former Prime Minister of Pakistan, Benazir Bhutto, found that police personnel had tampered with the case records [21]. Authorities of Bangladesh government systematically tamper or fabricate phone conversation and release it to social media with fake IDs, in order to frame or harass political opponents [22]

### 2.3 Withholding of Evidence

In 2016 governor of Bangladesh Bank (the central bank of Bangladesh) concealed what happened to be the largest bank heist anywhere in the world. Hackers managed to get into the SWIFT terminal computers of Bangladesh Bank and transferred around $1 billion from Bangladesh Bank's account with United States Federal Reserve to private accounts in Philippines and Sri Lanka [23]. When reported to the governor Dr. Atiur Rahman, instead of informing it to the authority and public, withheld information about the incident for about a month, until news about the incident appeared in Philippines' newspapers. Although external investigators (FBI, NSA) reported evidence of hackers from North Korea [24], internal report from Bangladesh government was never published because according to then finance minister of Bangladesh Abul Mal Muhit "influential names are on the report" [24]. In Pakistan, the JIT investigating Panama papers case declared that important documents regarding the source of income and money trail were withheld by the family of then prime minister, Mian Muhammad Nawaz Shareef [25]. The Federal Investigation Authority (FIA) of Pakistan reportedly face severe pressure to withhold information related to investigations against high profile cases [26]. Table 1 summarize few of these high profile cases from South Asian countries regarding evidence spoliation.

## 3 BLOCKCHAIN AND SMART CONTRACT

Distributed Ledger Technology (DLT), commonly known as blockchain, have demonstrable potential to prevent spoliation of evidence. Introduced just after the 2008 global financial crisis, blockchain is the underpinning technology of cryptocurrency Bitcoin [37]. Blockchain basically is an append-only, temper-evident, globally distributed digital ledger or database system where updated copy of the ledger is available to all stakeholders at all time and thereby provides transparency. Blockchain is also a "trustless" system where instead of trusted third party or intermediaries (e.g. banks, government organizations, corporations), trust on each transaction is validated by general consensus of the stakeholders through different consensus protocols [38]. Once validated, transaction is recorded on the ledger in an immutable fashion and the updated copy of the ledger is available to all stakeholders. Distributed and real time storage of consensus derived validated data, not only safeguard data from hacking but also tampering. Cryptographic capability of blockchain ensures privacy of the data producer.

Table 1: High Profile Cases of Spoliation of Evidence from South Asia

| Incident | Country | Summary | Category |
| --- | --- | --- | --- |
| JIT report on Lyari gang war leader Uzair Baloch | Pakistan | Government of Sind accused of tempering with the JIT report to protect high profile politicians on whose behest the gang war leader committed 198 (admitted) murders. Two versions of the report are circulating in media as of writing [27]. | Tampering / Withholding |
| Leaked call recordings by MP Ranjan Ramanayake | Sri Lanka | MP Hirunika Premachandra said that some of the recordings of telephone conversations between her and MP Ranjan Ramanayake had been tampered [19]. | Tampering |
| Chaudhry Suger Mills (Panama Papers) | Pakistan | Chairman Securities and Exchange Commission arrested for alleged tempering with financial records [28]. | Tampering |
| Income Tax Office, Bombay | India | Thousands of Suspicious Transaction Reports (STRs) and Dispute Resolution Receipts (DRRs) of the Income Tax Department were destroyed in fire [29]. | Destruction |
| Haryana High Court | India | Hundreds of court records were destroyed in fire. Report revealed that vital records were kept in fire prone rooms [30]. | Destruction |
| Bandarawela Court | Sri Lanka | Suspect with 11 pending cases set building on fire, with an objective of destroying evidence related to the case [31]. | Destruction |
| Deiyandara Court | Sri Lanka | Police suspect the fire that erupted at the evidence storage of the Deiyandara Magistrate's Court, was an act of arson [32]. | Destruction |
| Evidence destruction by Government's Chief Forensic Officer | Sri Lanka | Sri Lankan authorities on 7th August 2019 charged a former chief forensic officer with destroying evidence in the notorious murder of a leading rugby player [33]. | Destruction |
| Punjab Cooperative Bank Fire | Pakistan | Fire engulfed the whole floor where records of the housing societies investigated by authorities for corruption were kept [34]. | Destruction |
| Nandipur power project's Payment records | Pakistan | Records of multi-million rupees oil purchase destroyed in restricted access facility [35]. | Destruction |
| Bangladesh Bank Hack | Bangladesh | The Governor of the Bangladesh bank withheld the information about hack for $1 billion for over a month [24]. | Withholding |
| Asghar Khan Case | Pakistan | Rtd. Air Marshal urged the Supreme Court to investigate election rigging. Concerned authorities took six years to respond to the petition. Another five years passed due to reluctance of political parties [36]. | Withholding |

Recent years also witnessed the advent of smart contracts which are basically agreements written in computer codes. Instead of depending on third parties like solicitors to execute an agreement, smart contract automatically executes the agreed actions (e.g. payment) upon completion of a validated and recorded transaction on corresponding blockchain. Smart contract also enables data producer to take ownership of their data by implementing data sharing policies with data consumers [39]. For example, whom to share the data with, which part(s) of the data set and for how long.

## 4 RELATED WORK

Governments and academics have been actively exploring blockchain technology as a "source of truth" due to its intrinsic properties of provenance, persistence, auditability, traceability, transparency and immutability [40]. The state of Vermont of USA has legislated the blockchain based evidence submission [41]. The supreme people's court of China have embraced blockchain technology to tackle challenges in collection, preservation and authentication of evidence in the internet courts [42]. Her Majesty Court and Tribunals (HMCTS) of UK have been investigating the feasibility of using DLTs for securing digital evidence and evidence sharing with other law enforcement agencies [43]. Spring [9] argued that blockchain evidence is mostly inadmissible in the current Federal Rules of Evidence (FRE), even though it should be admissible in most cases, therefore urges Congress to consider amendments in FRE so that the "extraordinarily reliable" blockchain evidence is admissible.

Zhihong et al. [44] proposed a private blockchain system utilizing multi-signature solution for evidence submission and retrieval. The proposed solution named "Block-DEF" partially satisfies the traceability of evidence and treats evidence as a confidential document. However, since the system requires user to submit evidence to a private network, it does not guarantee the security or fairness of the system. Bonomi et al. [45] proposed a blockchain based chain of custody (BCOC) to ensure the traceability and integrity of the evidence and guaranteeing the state of evidence during the investigation. BCOC for the integrity and tamper resistance of digital forensics chain of custody has also been proposed by Lone and Mir [46], and Brotis et al.[47] propose a blockchain based solution for collection and preservation of forensic evidence in smart homes, using a permissioned blockchain. Jung et al. [48] proposed a framework for communication of Internet of Things (IoT) devices on public blockchain, in order to preserve the integrity of data and enhanced security, aiding transparent investigation [48].

Dimaz and Dony [49] highlighted a challenge in tax fraud investigation which relies on two copies of a letter as a proof of digital forensic acquisition. Either of the parties (taxpayer or tax investigator) can tamper with the letter, challenging the authenticity of the letter held by the counterparty. Case study conducted by Dimaz and Dony, demonstrates the usage of Bitcoin blockchain as a solution to prevent evidence tampering that may have occurred during a tax fraud investigation. Bela et al. [50] propose the usage of blockchain technology as a solution to guarantee the integrity of dashboard video footage so that it can be utilized as an evidence in court. Mingda et al. [51] proposed recording the hash of the campus surveillance video on to the "proof of stake" protocol based 'videochain' blockchain to ensure the credibility of the video evidence.

Most of the work done focuses on logging information on the blockchain for evidential purposes but to the best of our knowledge, no complete framework for evidence collection from whistle blowers/general public utilizing blockchain technology have yet been proposed. We explore not only the technical abilities of using a blockchain based solution for the protection of evidence spoilage but also give considerable attention to the privacy factor for both the evidence and evidence provider.

## 5 EVIDENCECHAIN: A PUBLIC-PRIVATE HYBRID BLOCKCHAIN MODEL

Any technical solution for preventing spoliation of evidence must meet the following conditions:

- Usability - Anyone should be able to come forward and provide evidence.
- Privacy - Evidence provider should be able to do so anonymously, since it is not only the evidence, but the evidence provider who also must be protected.
- Security – Guarantee that the evidence cannot be modified maliciously after submission and one could prove that the evidence in question is same as the one submitted by the evidence provider.

The proposed solution should permit members of public to provide evidence of public interest securely and anonymously, whilst simultaneously allowing the pertinent public authorities to utilize the system with escalated privileges. Therefore, we adopt a public-private hybrid blockchain model which allows to setup a consumer facing public blockchain (allowing everyone to participate) and a permissioned distributed ledger for legal cooperation and investigation proceedings within the authorities. We propose utilizing (a) a Tor [52] website for evidence provider's privacy, (b) distributed storage to avoid creating data honeypots which can be tampered or destroyed easily, (c) a public blockchain network to guarantee the open access, immutability and auditability of the evidence and (d) a private distributed ledger for secure collaboration among authorities and vetting of evidence. Architecturally, the model can be divided into three layers, (L1) *application layer* hosting the website, ensuring end user privacy and sanction resistant access, (L2) *storage layer* storing large evidence files in a distributed fashion, and (L3) *executive layer* with a hybrid distributed ledger structure to guarantee evidence immutability and secure data sharing within the investigating authorities.

### 5.1 Protecting privacy of evidence provider

Essentially, there can be two user groups within the system. 1) anonymous whistle blowers who would like to bring evidence to the light, 2) case workers and collaborators who wish to securely preserve the evidence. For user group 1, we take a *privacy by default* approach and propose using a Tor [52] website as the front-end for evidence submission in order to safeguard their identity and privacy. Tor is an anonymity network based on the *onion routing* introduced by Reed et al. [53]. It transmits data from source to destination using randomly selected route of nodes (a physical end device capable of receiving, sending, and forwarding information), utilizing asymmetric cryptography. The National Security Agency (NSA) termed Tor as "the King of highly secure, low latency Internet anonymity" [54]. Tor has become the *de-facto* communication software of choice in places where communication is monitored, and free speech is prohibited [52]. Investigative newspapers like The Guardian use Tor for secure drop service [55] and security expert like Edward Snowden who along with The Guardian journalists exposed the US National Security Agency (NSA) about their illegal global surveillance activities, thinks "Tor is the most important privacy-enhancing technology project being used today" [56]

Tor interface will allow evidence providers to submit evidence using average computers but leaving no digital footprint, having the confidence that the information or their identity cannot be compromised, and they cannot be traced or identified. Recent developments like Brave web browser have made access to Tor network very easy. With its 5 million daily active users, Brave can allow users to easily connect to Tor network without being isolated easily [57]. Privacy of the user can be further enhanced by using virtual private networks (VPN) or other anonymization techniques. However, the usability of the solution compels us to allow the user to upload evidence with minimum efforts.

In some cases, the evidence may contain privacy-sensitive information and the user may wish to not make it visible to public. The application layer would allow for the user to encrypt the evidence using the consortium's public keys on client side, before uploading to IPFS. The challenge with evidence encryption is that whistle-blower is trusting the consortium to act in all fairness (similar to [44]). However, it still gives the whistle blower a choice to release the evidence in public domain if they suspect collusion.

Group 2 users are usually members of the authorities and therefore can directly access the storage layer and communicate with the blockchain layer since they are privileged users who do not have anonymity concerns. They can store the encrypted evidence to the distributed storage, without having to mask their identity through Tor website. We recommend using attribute based [58] or hierarchical [59] encryption schemes for evidence encryption so that the evidence is not lost in case of user losing their private keys.

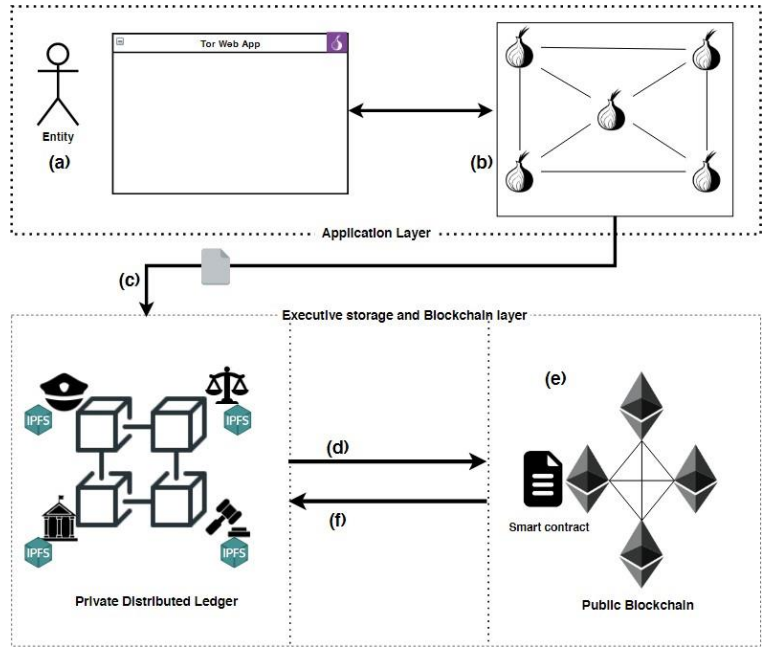

Figure 1: (a) The evidence provider uploads the digital evidence to the secure Tor website (solid lines) which is anonymized through Tor network (b). The content of the evidence is written to (c) private IPFS network of legal entities. Upon successful storage in IPFS, (d) the content-based hash from IPFS and metadata of the evidence is posted to Ethereum blockchain by invoking the evidence protection smart contract (e). Once the transaction has been accepted on the Ethereum blockchain (f) the metadata of the transaction/ evidence is posted back to the private distributed ledger, serving as an immutable record of the evidence.

### 5.2 Tamper evident distributed storage

As discussed in the section II, majority of the evidence tampering, and destruction occurs while evidence is stored at a centralized physical or virtual location. These evidence honeypots become the target for attackers to destroy or tamper with the evidence at rest. We propose to store the evidence in a private configuration of decentralized storage, IPFS [11], for each node in the evidence-chain consortium. This makes the evidence destruction a near impossible task since the evidence is not hosted at a single server or provider, instead it is replicated and distributed among several geographically distributed, decentralized nodes. Furthermore, IPFS addresses the files by content instead of the location, i.e. the cryptographic hash of the content becomes the address of the content. This not only guarantees the integrity of the evidence, since the cryptographic hash of evidence will change if a malicious actor tampers with the evidence, it permits to destroy the evidence permanently at the end of its life cycle. However, since we are relying on a private IPFS network in order to preserve the data privacy and confidentiality, the chances of collective tempering are higher as compared to using a public IPFS network.

### 5.3 Immutability of evidence

Evidence stored in IPFS is mutable and it can be modified. Therefore, to preserve the integrity and proof of upload, the IPFS resource locator (hash of content) along with its metadata is uploaded to a censorship resistant public blockchain like Ethereum, once the content is stored in IPFS. Public blockchain will guarantee the immutability and integrity of the commitment of evidence for perpetuity. The timestamping will also serve as a proof of existence for the evidence. Even though transacting on the public blockchain will incur extra costs, we believe that it is important for the trustworthiness and functioning of the system to be out of control of a single authority or a consortium of stakeholders with somewhat similar interests. For example, if the consortium collectively deletes or temper the evidence from all IPFS nodes, the hash on the Ethereum blockchain will point to nothing or will not match with the evidence, confirming that the evidence has been tempered with. However, the evidence can still be deleted at the end of its lifecycle. The smart contract will require signatures from all parties, confirming that they have removed the evidence following the legal proceedings such as court orders.

### 5.4 Collaboration and vetting of evidence

The judicial system works better if everyone can participate in evidence provision, especially if they have faith in the process.

However, utilizing the evidence in an investigation and using it in the court of law requires certain privileges, i.e. everyone cannot participate in these procedures. We propose a private consortium of authorities to vet the evidence and efficiently collaborate with each other on investigations.

Upon acceptance of the transaction on the public blockchain, the metadata of the transaction along with the metadata of the IPFS storage can be made available in the private distributed ledger network for verification by the relevant authorities. The authorities can collectively accept or reject the evidence on merit and admissibility. Humanitarian organizations and watchdogs can also be invited to the consortium, seeking fairness and transparency in the process. Fig. 1 shows a high-level view of the proposed model.

Fake evidence can be created and submitted as an evidence in order to sabotage one's reputation and can severely harm one's social position. Keeping personal data off the blockchain and only posting the commitment provides partly address the challenge since it still allows the user to remove the underlaying data. We also recommend liking the outcome of the investigation back to the evidence, so that observers are aware of the credibility of the evidence (similar to deletion of evidence as an outcome of investigation).

## 6  EVALUATION OF EVIDENCECHAIN

Strength of blockchain lies with its unique immutability, traceability and distributed features. Once information is archived in blockchain, the proof of evidence will remain for perpetuity in an immutable manner. Similar to internet, blockchain network cannot be taken down by any authority, let alone powerful individuals. No authorities can destroy, obstruct or even edit the content (cryptographic proofs of evidence) archived in public blockchain. This censorship resistant capability prevents evidence from attempt of spoliation, especially through destruction as exemplified in South Asian countries context.

Tampering of documents stored in distributed storage is also not possible because any change even a punctuation like semicolon, will break the hash that cryptographically represents the document on the blockchain. However, if tampered evidence were uploaded by member of public or authority itself, as in the cases of Sri Lanka, Pakistan and Bangladesh, that tampered evidence will remain on the network. In such scenario, member of the public or whistle blower within the authority in possession of the original/authentic evidence can anonymously publish it and make the visibility open to all. Depending on the public interest on the evidence, contradictory evidence may trigger public uproar in social media which may compel authorities to reinvestigate or even nullify the fake or tampered evidence, because of the existence of multiple versions of the evidence.

Officials authenticating fabricated or tampered evidence on the blockchain will have to think carefully. As the evidence authenticator, their names and signatures will be associated with the evidence and it can be retrieved easily using the traceability feature of blockchain and can be made publicly visible (like name and shame). Any future change of external or internal "power balance" or "mutual understanding" may jeopardize their career, entitlements, and social status, leading to potential prosecutions. However, since ordinary citizens can upload evidence anonymously, it is expected that in this model, many tampered/fabricated evidences will be made publicly available. Artificial Intelligence will play a pivotal role in this perspective to filter out tampered evidence, but collective intelligence and surveillance of ordinary citizen will ultimately enable to filter the right one [60].

Timestamping ability of blockchain can resists withholding of evidence. In Bangladesh Bank example, when governor of the bank instructed his subordinate within the bank to maintain "strict secrecy" just after the incident of heist was discovered, whistle blower within the bank could anonymously upload evidence on the *EvidenceChain* which would have been automatically timestamped. Whistle blower could subjectively give access of the evidence to public as when and how much required to demonstrate the level of evidence withheld. Evidence associated block timestamp would by default provide the length of withholding.

## 7  LIMITATIONS

The rising transaction fees and volatility are a major concern for the adaptation of blockchain technology and platforms like *EvidenceChain* which rely on the utilization of public blockchain. The average transaction fee (7 days simple moving average) on Ethereum network as of July 2020 is approximately USD 1.85, which is 10x increase from one year ago [61]. The transaction fees can also spike during the network congestion periods, as seen in the 2017-2018 cryptocurrencies boom (average transaction fees in the range of USD 3 per transaction). Transaction fees have spiked more recently to as high as $40 per transaction. One key limitation of our proposed model is transaction fee cost, i.e. who incurs the cost of evidence provision? A decentralized autonomous organization (DAO) funded by the public authorities, NGOs or individuals can be setup so that evidence providers can upload evidence for free. However, this may lead to a denial-of-service (DoS) attack by vested interest groups or funds being consumed by non-vital evidences submissions. Therefore, it is comparatively beneficial to have the evidence provider pay for the transaction. This will discourage the users of the system to submit fabricated or less useful transactions. The challenge can be somewhat addressed by linking the evidence to the user but we have carefully chosen anonymity over privacy.

Even though the proposed model intensively focuses on the evidence provider's privacy, security of the evidence provider cannot be fully guaranteed since the model is expected to be used by the public who are not always technology savvy. Onion routing service provided by Tor offers end to end encryption and robust security but since the domains are hashes of the public keys, they are difficult to remember and prone to typing errors, etc. Although the model guarantees the integrity of evidence once it has received sufficient verifications on the public blockchain, a malicious actor may temper with the evidence before the upload. Trusting oracles for real world data provision is an active area of research [62], however, no fully robust alternatives have been conceptualized yet. Furthermore, the source can be potentially identified based from the access means of evidence or the evidence itself, for example, a video footage capturing the evidence provider recording the evidence, hence revealing their identity.

## 8 FUTURE WORK

Similar to the *vanity onion domain*s of Facebook (<ext-link xmlns:xlink="http://www.w3.org/1999/xlink" ext-link-type="facebook" xlink:href="https://facebookcorewwwi.onion/">facebookcorewwwi.onion</ext-link>) or New York Times (nytimes3xbfgragh.onion), we aim to generate onion addresses as variations of keyword 'evidence' to improve on the accessibility of an onion websites for evidence protection and increase the identity protection for the evidence provider. We also plan to develop a working model of the proposed system so that the practical aspects can be evaluated. In future, we would revisit the proposed model to adapt for decentralized blockchain domains, once the eco system of decentralized web has evolved in the years to come.

## 9 CONCLUSION

Although the effectiveness of the model to prevent spoliation of evidence is theoretically and technically achievable, the success of the model lies with usability, particularly how simple the experience is for the general public to upload evidence to the platform anonymously. If users cannot access the platform easily, it will hinder the wider usage and effectiveness of the system. If the web interface is available via the open web, then it would compromise user's identity and authorities could also monitor the platform to target the users. Therefore, a Tor web service-based interface is proposed by this model which although has robust capability to maintain privacy but in terms of usability, non-technical users may find it difficult, especially in contrast to open web interface. Success of the model therefore lies with finding the right balance between usability and privacy, a paradox faced by all digital applications [63].

Transforming present centralized, tamper and attack prone evidence management system to a distributed, immutable and indestructible system through *EvidenceChain* like model will have two societal impacts: First, empowering ordinary citizen and whistle blower with a new paradigm, where they will be able to exert their power to prevent spoliation of evidence as well as human rights abuse by publishing evidence without the fear of exposure and retribution. Second, transforming the vetting process into a consensus-based approach will prevent any compromised actor within the authorities to manipulate the vetting process. Moreover, transparency, especially linking vetted evidence with signatories and preserving the integrity record immutably on public blockchain, will transfer the fear of retribution from ordinary citizen to authorities. In South Asian and other developing countries context, this transfer of power from authority to public and fear of retribution other way round, may contribute to change the post-colonial age mindset, which we believe in long run will not only minimize the spoliation of evidence but may maximize the potential to establish transparent and just societies.

## REFERENCES
[1] H. Campbell Black and M. A. Author, "Black's law dictionary," 1999.
[2] J. Rogers, "Prosecutorial Crimes and Corruption: The (White) Elephant in the Courtroom," 2017.
[3] J. Cohen, "Studies on Transitional Justice in Context Addressing Corruption Through Justice-Sensitive Security Sector Reform," 2017.
[4] T. Wyatt, K. Johnson, L. Hunter, R. George, and R. Gunter, "Corruption and Wildlife Trafficking: Three Case Studies Involving Asia," Asian J. Criminol., vol. 13, no. 1, pp. 35–55, Mar. 2018.
[5] A. Kouznetsov, S. Kim, and J. Pierce, "A longitudinal meta-analysis of corruption in international business and trade: Bridging the isolated themes," Int. Trade J., vol. 32, no. 5, pp. 414–438, Oct. 2018.
[6] K. A. Emmanuel, "Effect of Corruption on Corporate Governance in Selected Area Offices of Deposit Money Banks in Enugu State, Nigeria," Int. Res. J. Manag. IT Soc. Sci., vol. 4, no. 2, 2017.
[7] M. A. Zuckerman, "YES, I DESTROYED THE EVIDENCE-SUE ME? INTENTIONAL SPOLIATION OF EVIDENCE IN ILLINOIS."
[8] J. Warkotsch *et al.*, Beyond the Panama Papers. The Performance of EU Good Governance Promotion: The Anticorruption Report, volume 4, 1st ed. Verlag Barbara Budrich, 2017.
[9] J. C. Spring, "The Blockchain Paradox: Almost Always Reliable, Almost Never The Blockchain Paradox: Almost Always Reliable, Almost Never Admissible Admissible THE BLOCKCHAIN PARADOX: ALMOST ALWAYS RELIABLE, ALMOST NEVER ADMISSIBLE," SMU Law Rev., vol. 72, no. 4, 2019.
[10] V. Buterin, "A next-generation smart contract and decentralized application platform," Etherum, no. January, pp. 1–36, 2014.
[11] J. Benet, "IPFS - Content Addressed, Versioned, P2P File System," Jul. 2014.
[12] "Tor Project | Anonymity Online." [Online]. Available: https://www.torproject.org/. [Accessed: 02-Jun-2020].
[13] X. Xu *et al.*, "A taxonomy of blockchain-based systems for architecture design," Ieeexplore.Ieee.Org.
[14] "LAHORE: Fire breaks out in LDA building - Newspaper - DAWN.COM." [Online]. Available: https://www.dawn.com/news/361501. [Accessed: 22-May-2020].
[15] "Eight plunge to death as fire erupts at LDA plaza in Lahore | The Express Tribune." [Online]. Available: https://tribune.com.pk/story/546486/several-trapped-as-fire-erupts-at-lda-plaza-in-lahore/. [Accessed: 22-May-2020].
[16] "PNB fraud: Nirav Modi papers gutted in fire at Income Tax office: The Tribune India." [Online]. Available: https://www.tribuneindia.com/news/archive/nation/pnb-fraud-nirav-modi-papers-gutted-in-fire-at-income-tax-office-599478. [Accessed: 04-Jun-2020].
[17] "BSTI destroys evidence, frames innocents: court." [Online]. Available: https://www.newagebd.net/article/97234/bsti-destroys-evidence-frames-innocents-cou. [Accessed: 02-Jun-2020].
[18] "Sri Lanka ex-minister held over phone calls recording scandal- The New Indian Express." [Online]. Available: https://www.newindianexpress.com/world/2020/jan/17/sri-lanka-ex-minister-held-over-phone-calls-recording-scandal-2090807.html. [Accessed: 02-Jun-2020].
[19] S. Indrajith, "The Island - Hirunika says her conversations with Ranjan tampered with." [Online]. Available: http://island.lk/index.php?page_cat=article-details&page=article-details&code_title=217537]. I. [Accessed: 04-Jun-2020].
[20] "Sahiwal killings forensic report finds CTD tampered with digital evidence." [Online]. Available: https://www.geo.tv/latest/228221-sahiwal-killings-forensic-report-finds-ctd-tampered-with-digital-evidence. [Accessed: 22-May-2020].
[21] "Benazir Bhutto assassination case: JIT complete report | The Express Tribune." [Online]. Available: https://tribune.com.pk/story/339661/benazir-bhutto-assassination-case-jit-complete-report/. [Accessed: 22-May-2020].
[22] "ফোনালাপ ফাঁস, যা বললেন ডিপি নুর | 846890 | কালের কণ্ঠ | kalerkantho." [Online]. Available: https://www.kalerkantho.com/online/national/2019/12/04/846890. [Accessed: 02-Jun-2020].
[23] "The Billion-Dollar Bank Job - The New York Times." [Online]. Available: https://www.nytimes.com/interactive/2018/05/03/magazine/money-issue-bangladesh-billion-dollar-bank-heist.html. [Accessed: 28-May-2020].
[24] "ফের কথা পাল্টালেন অর্থমন্ত্রী | daily nayadiganta." [Online]. Available: https://www.dailynayadiganta.com/detail/news/155682?m=0. [Accessed: 28-May-2020].
[25] W. Zia, B. Rasul, K. Khurshid, A. Aziz, N. Saeed, and I. Mangi, "FINAL INVESTIGATION REPORT OF JOINT INVESTIGATION TEAM-(PANAMA CASE) The Scope and Key Focus Areas of JIT's Investigation Report."
[26] U. Cheema, "Efforts on to save accused." [Online]. Available: https://nation.com.pk/05-Dec-2009/efforts-on-to-save-accused. [Accessed: 04-Jun-2020].
[27] "Uzair Baloch JIT report: Ali Zaidi appeals to CJP to take suo motu notice." [Online]. Available: https://www.thenews.com.pk/latest/683414-uzair-baloch-jit-report-ali-zaidi-appeals-to-cjp-to-take-suo-motu-notice. [Accessed: 09-Jul-2020].
[28] "Record tempering case: Court orders trial of ex-SECP chief Zafar Hijazi - SAMAA." [Online]. Available: https://www.samaa.tv/news/2017/11/record-tempering-case-court-orders-trial-ex-secp-chief-zafar-hijazi/. [Accessed: 05-Jun-2020].
[29] "Nirav Modi files shifted out before blaze: I-T probe - The Hindu." [Online]. Available: https://www.thehindu.com/news/cities/mumbai/nirav-modi-files-shifted-out-before-blaze-i-t-probe/article24436986.ece. [Accessed: 02-Jun-2020].
[30] "Court records destroyed in fire | Chandigarh News - Times of India." [Online]. Available: https://timesofindia.indiatimes.com/city/chandigarh/Court-records-destroyed-in-fire/articleshow/7393788.cms. [Accessed: 02-Jun-2020].
[31] "Bandarawela court arsonist arrested | Daily News." [Online]. Available: https://www.dailynews.lk/2018/05/09/law-order/150398/bandarawela-court-arsonist-arrested. [Accessed: 02-Jun-2020].

[32] "Fire at Deiyandara Court suspected to be an act of Arson to destroy evidence." [Online]. Available: https://www.newsfirst.lk/2020/05/14/fire-at-deiyandara-court-suspected-to-be-an-act-of-arson-to-destroy-evidence/. [Accessed: 04-Jun-2020].

[33] "Sri Lanka Charges Forensic Chief over Rugby Murder Cover Up, Could Get Death Sentence." [Online]. Available: https://www.news18.com/news/world/sri-lanka-charges-forensic-chief-over-rugby-murder-cover-up-could-get-death-sentence-2262181.html. [Accessed: 04-Jun-2020].

[34] "Ashiana, Paragon housing scheme records go up in flames in Lahore - SAMAA." [Online]. Available: https://www.samaa.tv/news/2018/08/ashiana-paragon-housing-scheme-records-go-up-in-flames-in-lahore/. [Accessed: 05-Jun-2020].

[35] "Nandipur power project payments, purchases record gutted." [Online]. Available: https://www.thenews.com.pk/print/148967-Nandipur-power-project-payments-purchases-record-gutted. [Accessed: 05-Jun-2020].

[36] "A story behind Asghar Khan case?" [Online]. Available: https://www.geo.tv/latest/195530-a-story-behind-asghar-khan-case. [Accessed: 06-Jun-2020].

[37] S. Nakamoto, "Bitcoin: A Peer-to-Peer Electronic Cash System," Www.Bitcoin.Org, p. 9, 2008.

[38] A. Shahaab, B. Lidgey, C. Hewage, and I. Khan, "Applicability and Appropriateness of Distributed Ledgers Consensus Protocols in Public and Private Sectors: A Systematic Review," IEEE Access, vol. 7, pp. 43622–43636, 2019.

[39] G. Zyskind, O. Nathan, and A. S. Pentland, "Decentralizing privacy: Using blockchain to protect personal data," Proc. - 2015 IEEE Secur. Priv. Work. SPW 2015, pp. 180–184, 2015.

[40] A. Shahaab, R. Maude, C. Hewage, and I. Khan, "Blockchain: A Panacea for Trust Challenges In Public Services? A Socio-technical Perspective," vol. 3, no. 2, pp. 1–11, 2020.

[41] "Vermont Laws." [Online]. Available: https://legislature.vermont.gov/statutes/section/12/081/01913. [Accessed: 29-May-2020].

[42] "Provisions of the Supreme People 's Court on Several Issues Concerning the Trial of Cases by Internet Courts-Supreme People 's Court of the People 's Republic of China." [Online]. Available: http://www.court.gov.cn/zixun-xiangqing-116981.html. [Accessed: 29-May-2020].

[43] "How we're investigating Digital Ledger Technologies to secure digital evidence - Inside HMCTS." [Online]. Available: https://insidehmcts.blog.gov.uk/2018/08/23/how-were-investigating-digital-ledger-technologies-to-secure-digital-evidence/. [Accessed: 21-May-2020].

[44] Z. Tian, M. Li, M. Qiu, Y. Sun, and S. Su, "Block-DEF: A secure digital evidence framework using blockchain," Inf. Sci. (Ny)., vol. 491, pp. 151–165, Jul. 2019.

[45] S. Bonomi, M. Casini, and C. Ciccotelli, "B-CoC: A blockchain-based chain of custody for evidences management in digital forensics," in OpenAccess Series in Informatics, 2020, vol. 71.

[46] A. Hamid Lone and R. Naaz Mir, "FORENSIC-CHAIN: ETHEREUM BLOCKCHAIN BASED DIGITAL FORENSICS CHAIN OF CUSTODY," 2017.

[47] S. Brotsis et al., "Blockchain Solutions for Forensic Evidence Preservation in IoT Environments."

[48] J. H. Ryu, P. K. Sharma, J. H. Jo, and J. H. Park, "A blockchain-based decentralized efficient investigation framework for IoT digital forensics," J. Supercomput., vol. 75, no. 8, pp. 4372–4387, Aug. 2019.

[49] D. Ankaa Wijaya and D. Ariadi Suwarsono, "Securing Digital Evidence Information in Bitcoin A Case Study in Directorate General of Taxes."

[50] B. Gipp, J. Kosti, and C. Breitinger, "Securing Video Integrity Using Decentralized Trusted Timestamping on the Bitcoin Blockchain," MCIS 2016 Proc., Jan. 2016.

[51] M. Liu, J. Shang, P. Liu, Y. Shi, and M. Wang, "VideoChain: Trusted Video Surveillance Based on Blockchain for Campus," in Lecture Notes in Computer Science (including subseries Lecture Notes in Artificial Intelligence and Lecture Notes in Bioinformatics), 2018, vol. 11066 LNCS, pp. 48–58.

[52] R. A. Haraty and B. Zantout, "The TOR Data Communication System: A Survey."

[53] M. G. Reed, P. F. Syverson, and D. M. Goldschlag, "Anonymous Connections and Onion Routing."

[54] "Tor: 'The king of high-secure, low-latency anonymity' | US news | theguardian.com." [Online]. Available: https://www.theguardian.com/world/interactive/2013/oct/04/tor-high-secure-internet-anonymity. [Accessed: 27-May-2020].

[55] "The Guardian SecureDrop." [Online]. Available: https://www.theguardian.com/securedrop. [Accessed: 14-Jul-2020].

[56] "Edward Snowden Explains How To Reclaim Your Privacy." [Online]. Available: https://theintercept.com/2015/11/12/edward-snowden-explains-how-to-reclaim-your-privacy/. [Accessed: 14-Jul-2020].

[57] "Brave passes 15 million monthly active users and 5 million daily active users, showing 2.25x MAU growth in the past year." [Online]. Available: https://brave.com/15-million/. [Accessed: 09-Jul-2020].

[58] J. Bethencourt, A. Sahai, and B. Waters, "Ciphertext-Policy Attribute-Based Encryption," 2007.

[59] M. Abdalla, E. Kiltz, and G. Neven, "Generalized key delegation for hierarchical identity-based encryption," in European Symposium on Research in Computer Security, 2007, pp. 139–154.

[60] N. Stephens, I. Khan, and R. Errington, "Analysing the role of virtualisation and visualisation on interdisciplinary knowledge exchange in stem cell research processes," Palgrave Commun., vol. 4, no. 1, Dec. 2018.

[61] "Ethereum Avg. Transaction Fee chart." [Online]. Available: https://bitinfocharts.com/comparison/transactionfees-eth-sma7.html. [Accessed: 09-Jul-2020].

[62] A. Egberts, "The Oracle Problem - An Analysis of how Blockchain Oracles Undermine the Advantages of Decentralized Ledger Systems," SSRN Electron. J., Jun. 2019.

[63] "Security vs usability: it doesn't have to be a trade-off." [Online]. Available: https://www.telegraph.co.uk/connect/better-business/security-versus-usability-ux-debate/. [Accessed: 02-Jun-2020].